\documentclass[conference]{IEEEtran}

\usepackage{amsmath,amssymb,amsfonts}
\usepackage[linesnumbered,ruled]{algorithm2e}
\usepackage{graphicx,xcolor,array}
\usepackage{tikz}
\usetikzlibrary{arrows}
\usetikzlibrary{automata,positioning}
\usepackage{makecell}
\usepackage{upgreek}
\usepackage{hyperref}
\usepackage{url}

\newcommand{\shorten}[1]{}
\usepackage{array}
\newcolumntype{L}[1]{>{\raggedright\let\newline\\\arraybackslash\hspace{0pt}}m{#1}}
\newcolumntype{C}[1]{>{\centering\let\newline\\\arraybackslash\hspace{0pt}}m{#1}}
\newcolumntype{R}[1]{>{\raggedleft\let\newline\\\arraybackslash\hspace{0pt}}m{#1}}
\usepackage{multirow}
\usepackage{textcomp}
\usepackage{xcolor}

\begin{document}

\title{GDPR Compliance for Blockchain Applications in Healthcare}

\author{ 
\author{
	\IEEEauthorblockN{Anton Hasselgren\IEEEauthorrefmark{1}, Paul Kengfai Wan\IEEEauthorrefmark{2}, Margareth Horn\IEEEauthorrefmark{3}, Katina Kralevska\IEEEauthorrefmark{4}, Danilo Gligoroski\IEEEauthorrefmark{4} and Arild Faxvaag\IEEEauthorrefmark{1} \\ 
	\IEEEauthorblockA{\IEEEauthorrefmark{1}Dep. of Neuromedicine and Movement Science, NTNU, Norwegian University of Science and Technology}
	\IEEEauthorblockA{\IEEEauthorrefmark{2}Dep. of Manufacturing and Civil Engineering, NTNU, Norwegian University of Science and Technology}
	\IEEEauthorblockA{\IEEEauthorrefmark{3}Dep. of Social- and Educational Science, NTNU, Norwegian University of Science and Technology}
	\IEEEauthorblockA{\IEEEauthorrefmark{4}Dep. of Information Security and Communication Technology, NTNU, Norwegian University of Science and Technology}
	Email: \{anton.hasselgren, paul.k.wan, margareth.horn, katinak, danilog, arild.faxvaag\}@ntnu.no}
}

}

\maketitle

\begin{abstract}
The transparent and decentralized characteristics associated with blockchain can be both appealing and problematic when applied to a healthcare use-case. As health data is highly sensitive, it is also highly regulated to ensure the privacy of patients. At the same time, access to health data and interoperability is in high demand. Regulatory frameworks such as GDPR and HIPAA are, amongst other objectives, meant to contribute to mitigating the risk of privacy violations in health data. Blockchain features can likely improve interoperability and access control to health data, and at the same time, preserve or even increase, the privacy of patients. Blockchain applications should address compliance with the current regulatory framework to increase real-world feasibility. This exploratory work indicates that published proof-of-concepts in the health domain comply with GDRP, to an extent. Blockchain developers need to make design choices to be compliant with GDPR since currently, none available blockchain platform can show compliance out of the box.
\end{abstract}

\begin{IEEEkeywords}
Blockchain, DTL, health data, GDPR, privacy regulations
\end{IEEEkeywords}

\section{Introduction}
The current status in data privacy could be categorized as the post-privacy area due to the unintended consequences of the big data revolution. The famous Cambridge Analytica scandal \cite{berghel2018malice} is an example of how re-identification can be achieved by cross-analyze of large data sets containing private information. The technology revolution that has driven us to post-privacy has not been stopped through privacy acts such as General Data Protection Regulation (GDPR). We are currently in another (r)evolution that can restore data privacy - blockchain. 

In 2018, The European Union instituted the GDPR \cite{EUdataregulations2018}, which regulates the collection, processing and securing of personal data, including protected health information (PHI). Art. 4(15) of the EU GDPR, defines data concerning health as: “personal data related to the physical or mental health of a natural person, including the provision of healthcare services, which reveal information about his or her health status.” 

Health Insurance Portability and Accountability Act (HIPPA) is essential for U.S. healthcare law and deals mainly with privacy rights and access rather than ownership of patient data \cite{annas2003hipaa}. Each state has ownership of patient data of its citizens, and it is controlled by respective state law, in this case. Since there are 50 states, there are 50 differing laws, court cases and interpretations of that ownership of patient data. New Hampshire has as the only state enacted legislation stating that patient data ownership lies with the patient. How GDPR will interact and comply with U.S. state laws remains to be determined.

Blockchain, first introduced with the launch of Bitcoin back in 2008 has become more diluted in its definition, and there is currently no fixed, widely accepted definition of the term blockchain. To clarify its use in this research we have defined Blockchain as a distributed, de-centralized and tamper-proof ledger without any centralized control. Blockchain technology and other Distributed Ledger Technologies (DLT) could increase our data privacy and empower individuals with control and access over their data, including health data.

The objective of this study is twofold: (i) to dissect the various designs of blockchain and explore GDPR compliance for different components in established or proposed blockchain applications in the healthcare sector and, (ii) to provide a future researcher with guidance in how to comply with GDPR when designing blockchain application within the healthcare domain.

The rest of this paper is organized as follows: \textbf{Section II} provides a brief introduction to blockchain technology and outlines previous work addressing blockchain compliance with GDPR; \textbf{Section III} presents the research approach; \textbf{Section IV} describes four blockchain applications in healthcare identified through the literature; \textbf{Section V} presents our results and analysis; \textbf{Section VI} provides a discussion and conclusion to the work, and gives a recommendations for future work.

\section{GDPR and Blockchain}
This section gives a brief introduction to blockchain and GDPR. We can broadly categorize blockchain as; public permissionless, private permissioned and federated permissioned.  The categorization is important in order to design applications in relevant sectors to achieve social and economic goals. In a public blockchain, everyone in the network holds equal rights and the ability to access the ledger. While nodes need to be certified to join the consensus process in private and federated, which makes them permissioned. The French National Commission on Informatics and Liberty (CNIL) recommends private permissioned blockchains because of the compliance with GDPR \cite{EUblockchainobservatoryandforum}. The third blockchain category is a combination between public and private, which are referred to as a federated or hybrid blockchain \cite{casino2019systematic}. 

Previous research has explored blockchain platforms and their feasibility for healthcare \cite{kuo2019comparison} and concluded that none of the most widely used blockchain platforms were ideal for healthcare, out of the box. Other work has identified important properties and characteristics in different types of blockchain, and they need to be considered in the initial design phase; identity management, efficiency according to energy use, immutability, ownership management and transaction approval \cite{casino2019systematic}. Public, private and federated blockchains handle these properties differently and each component may have their limitation \cite{tasca2017taxonomy}.

As detailed by the European Union Blockchain Observatory and Forum \cite{EUblockchainobservatoryandforum}, in principle, there are no contradictions between the goals of GDPR and DLT. However, there seem to be at least three areas in which GDPR still does not offer enough clarity about how real-world DLT applications for the health sector should be developed: (1) accountability and roles (e.g., how to identify a data controller in a public DLT), (2) anonymization of personal data (e.g., which techniques are sufficient to anonymize personal data to the point where the resulting output can potentially be stored in a DLT), and (3) GDPR rights conflicts (e.g., how to rectify or remove personal data that are recorded in a DLT that is immutable by nature, or who is responsible for requesting and managing the “freely, specific, informed, and unambiguous” consent from a data subject, especially if the data controller is not specified) \cite{regulation2016regulation}. With regards to the anonymization of personal data, it is clear that GDPR does not apply to anonymized data and that this type of information can be stored on the open ledger. However, what qualifies as anonymized is still not clear. The only indication today is that it must be irreversibly impossible to identify an individual through any of the means “reasonably" likely to be used \cite{el2015critical}.

Smart contracts are one of the components that have been proposed on blockchain platforms to reduce the need for a third party. It enables a new type of autonomous regulation that executes transactions when all the requirements are fulfilled \cite{kim2017perspective}. All legal rules and contracts are transposed into digital and software rules, which means smart contract is the new regulator in the blockchain network and rules are followed accordingly \cite{de2018blockchain}. For example, after a user authenticates its digital identity successfully, smart contract can grant authorization and access his/her accurate medical records to the requestors \cite{wan2020blockchain}. However, Giordanengo \cite{giordanengo2019possible} analyzed some use cases of smart contracts and found out that none of the studies have reached the stage of production and concluded that it is not ready for implementation in healthcare domain. 

The design options in a blockchain application are wide, and there is an increase research and innovation in this field. To design blockchain applications for healthcare use-cases, there are several important design choices the developer has to make, with the three most prominent: (1) choice of platform/network, (2) on/off chain data storage and (3) identity solution for interaction with the system. 

\subsection{Related work}
This section highlights previous work which has investigated blockchain compliance with GDPR. There is limited published research under this topic in the literature, but previous work has indicated the need for standardization \cite{kuo2019comparison}, \cite{tasca2017taxonomy}, \cite{hasselgren2020gdpr}. 

Two reports published by EU entities: The EU blockchain Observatory and Forum  - Blockchain and the GDPR \cite{EUblockchainobservatoryandforum} and the European Parliamentary Research Service (EPRS) - Blockchain and the general data protection regulation \cite{EPRS2019} provide guidance in blockchain compliance with GDPR. 

Blockchain and the GDPR is a thematic report published in 2018 where accountability and roles, as well as anonymization of personal data, are addressed. This report highlights the need for each blockchain use-case to be thoroughly analyzed and rated in various interpretations - compliance with GDPR is not about the technology but rather, how it is used. The report also points out the need to avoid storing personal data on a public blockchain and anonymous data techniques such as obfuscation, encryption and aggregation should be used. The report proposed some principles to consider when designing blockchain architectures such as; considering user perspective, analysing where the personal data appears and who is responsible for the processing.

Blockchain and the GDPR report has defined roles for three main actors: data subject, data controller and data processor, and it is outlined in the report that it can be problematic to identify the data controller in blockchains. The report also presents some techniques, such as  reversible encryption and hashing, to achieve anonymous or pseudo-anonymous data, which is under intense debates. It is also important to consider if personal data should be involved when linking private chains with public chains.

In the report by the European Parliamentary Research Service, blockchain is defined as a combination of many different forms of distributed databases that present variation, both in complexity and governance agreement. The report gives an account of difficulties in whether personal data, can be anonymized to the extant that it meet the GDPR threshold of anonymization. Two types of compliance tension is expressed: (1) GDPR assumes that there is a data controller, which often is not the case in blockchains and (2) the right to be forgotten, which is problematic in an immutable ledger. Taking into evaluation some properties in the infrastructure level on blockchains, it is suggested possible controllership and reflect on the nexus between responsibility and control. Furthermore, it is expressed that it is difficult to assess the compatibility between blockchain and GDPR without having to pay attention to the nuances in blockchain configurations. There is also a request for further clarifications of concepts such as "anonymous data", "data controller" and the meaning of "erasure" under Art. 17. 

There are some examples where blockchains compliance with GDPR is tested, such as a proof-of-concept (PoC) developed by Hawig et al. \cite{hawig2019designing} for the use case of blood glucose data. They examined a system for immutable, interoperable and GDPR compliant data exchange. In this PoC, they highlight that blockchain has a great potential to improve information transactions in a secure and transparent way between patients and providers \cite{hawig2019designing}. They tested two possible solutions based on the public IOTA blockchain and in combination with public IOTA plus, a private IPFS (InterPlanetary File System) cluster. In the public IOTA it became difficult to eliminate the risk of personal data linkability, and combining a public DLT and IPFS has a high degree of complexity. They also highlight that there are limitations in identifying a data controller since the public DLT ecosystem is formed by multiple health care stakeholders, as well as patient consent management. They argue that each use-case must be carefully considered when blockchain-based system is designed for health data exchange.  
 
One private blockchain is suggested in the CUREX project, which is argued to have GDPR compliance by design in a decentralized architecture \cite{diaz2019overview}. In this project, they argue that all data transactions in health sectors and their vulnerability depend on private blockchain infrastructure to integrity of risk management. The CUREX project's goal is to ensure the integrity of the risk assessment process of all data transactions.
 
Two other suggested GDPR-compliant design concepts that address health data collected  by sensors in different types of mobile and smart devices. Both designs were described to address the vulnerability in centralised data storage controlled by service providers, and the "right to own and share personal information". One of the concepts is combining blockchain with cloud storage and machine learning techniques to give users the possibility to share personal data easily and securely. In this model, the data is encrypted before uploading it to the cloud storage and secured by the hash function. The access to the data is split and distributed among multiple key keepers, and no visible personal information is involved because the blockchain allows pseudonyms \cite{zheng2018blockchain}. While Zheng et al. considered the limits blockchain has to store large-size data that are continuous-dynamic, Hossein et al. propose architecture for efficient access and control mechanisms \cite{hossein2019blockchain}. In their work, the privacy challenge was addressed, and their design concept is an efficient privacy-preserving access to give the users full control over their own data. 

\section{Method}
The research approach in this work has been: (i) review of four different blockchain proof-of-concepts for healthcare; (ii) review of GDPR and how the regulations apply to health informatics; and (iii) an exploratory analysis how the platforms reviewed in (i) comply with the relevant articles identified in (ii).

The blockchain based proof-of-concepts in (i) were identified through a scoping search in PubMed and Scopus. The following documents were identified and utilized in (ii):

\begin{itemize}
    \item GDPR (official document) \cite{EUdataregulations2018};
    \item Blockchain and the General Data Protection Regulation \cite{finck2019blockchain}.
\end{itemize}

\section{Blockchain-based healthcare applications}
This section describes four different blockchain applications that were identified in a scoping search in PubMed and scopus. These four applications were included based on their different architecture. 
MedRec \cite{azaria2016medrec} is a blockchain-based solution for personal control of identity and the distribution of health information. The system is designed on the public Ethereum blockchain. This means that transactions, including metadata which is sensitive in medical context, are visible to everyone who has access to the blockchain. And if someone can identify the  patient's real world identity and Ethereum account, one can determine the relationship between the health providers and the patients. In order to circumvent this privacy issue, MedRec anonymized metadata through disassociating each patient's identity from the provider, where each provider makes a new identity Ethereum account for each patient-provider relationship. The purpose is to enable patients to establish public relations without revealing the real-world identities.

EMRshare  \cite{xiao2018emrshare} is a health data sharing application where different entities such as health provider, data scientists and patients interact using the permissioned Hyperledger blockchain. Transactions such as health data requests, approval or rejection action is stored on the blockchain. While actual medical data are stored off-chain and encrypted with asymmetric encryption for security purposes. EMRShare also enables patients, the data owner, to anonymize their name or identity in the medical records before reaching the requestors.

\begin{figure}
    \vspace{0.2cm}
    \centering
    \includegraphics[width=0.9\columnwidth]{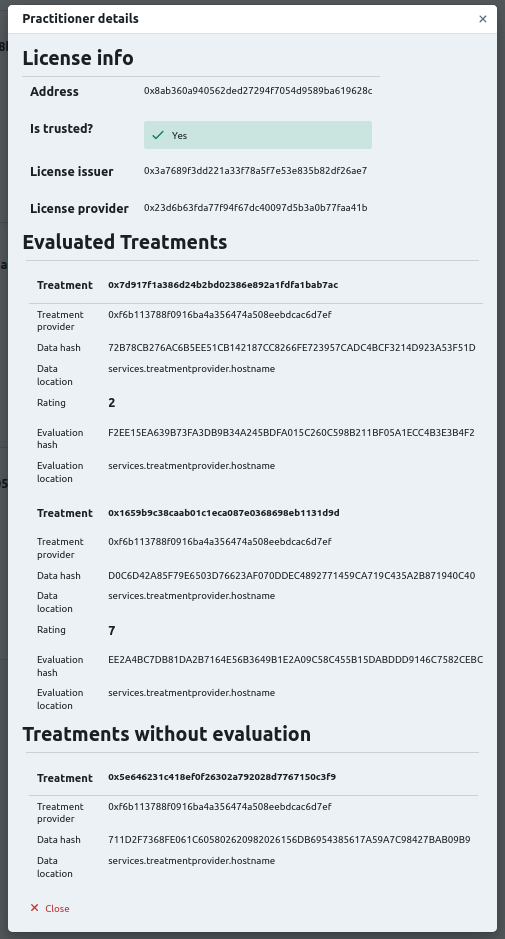}
    \caption{Page from the User Interface of VerifyMed for showing details about all data related to a healthcare worker \cite{JarensaaGithubTransparentHealthcare,Rensaa2020}.}
    \label{b:fig:proof-overview-details}
\end{figure}

VerifyMed \cite{10.1145/3409934.3409946,JarensaaGithubTransparentHealthcare,Rensaa2020} is a public Ethereum blockchain platform with the aim to validate the authorization and competence of healthcare workers in a virtualized healthcare environment. VerifyMed enables healthcare workers to document their work history and competence in the form of a de-centralized portfolio. VerifyMed combines and stores three forms of data items (evidence of authority, evidence of experience and evidence of competence) to build their portfolios. Digital signatures scheme is also incorporated in VerifyMed to ensure ownership is established on each verified evidence. As an example of how a typical user interface looks like in a blockchain-based application, we give Figure \ref{b:fig:proof-overview-details}, which is taken from VerifyMed\cite{JarensaaGithubTransparentHealthcare,Rensaa2020}.

FHIRChain \cite{zhang2018fhirchain}) is a public Ethereum blockchain architecture for secure and scalable clinical data sharing with the goal to meet the requirements of The Office of the National Coordinator for the Health Information Technology (ONC) such as privacy preserving and health information security. FHIRchain stores encrypted metadata on the network rather than storing encrypted sensitive health data. It uses digital health identity, which utilizes public-key cryptography to generate and manage the identities. Often clinical data research data format and structures varies from institution to institution, which makes data sharing challenging, FHIRChain is developed based on HL7 Fast Healthcare Interoperability Resources (FHIR) to enforce consistent data formats for easier information sharing. 

These proposed blockchain concepts within the healthcare domain primarily focus on solving interoperability without compromising the privacy and security of sensitive health data. Identity management for both patients and health workers is also considered as part of the proposed applications. However, the research work focusing on the degree of compliance  to GDPR and other health data regulatory frameworks remains limited.

\section{Results}
The results are presented mainly in Table \ref{table1} - relevance of GDRP for healthcare and analysis of compliance and Table II - Comparative analyses.

GDPR will have a significant impact on the healthcare sector in collecting, processing, and securing protected health information. Healthcare institutions (HI), are obligated to ensure that data is collected for a specific and legitimate use and that the data is only used for that purpose. Further,  a healthcare organization will be required to obtain exclusive consent or permission from the patient (the data subject) to use their data according to (Art. 7). 

GDPR indicates that the ownership of health data should be with the patients, enabling patients to have greater autonomy over their data. Healthcare providers are obligated to furnish patients with complete information when they request it, within specified time limits (Art. 15). 

Additionally, GDPR requires organizations to report a data breach within 72 hours of learning of the breach (Art. 33), and notify the affected individual if the breach results in an adverse impact (Art. 34). The onus will, therefore, be on healthcare organizations to ensure that data is highly secured and protected from unauthorized access or face rapid reporting requirements of breaches and possible severe financial penalties (see Art. 83). This is important when it comes embracing new digital technologies like blockchain because issues such as misuse of patients' PHI would result in losing trust in healthcare institutions and delaying the adoption of blockchain in enhancing information sharing.

Table I summarizes what healthcare institutions (HI) must consider when using blockchain and DLT to secure and protect PHI and avoid costly fines for non-compliance. 

\begin{table*}[h!]
\centering
    \caption{Relevance of GDPR for healthcare}\label{table1}
        \begin{center}
        \begin{tabular}{|L{4cm}|L{4cm}|L{6cm}|}
        \noalign{\hrule height 2pt}
     
        \textbf{Article in GDPR} & \textbf{{Compliance}}& \textbf{{Impact in healthcare}} \\
        \noalign{\hrule height 2pt}
        \textbf{Art. 30 (Records of processing activities),
        Art. 35 (Data protection impact assessment)
        } & {{Able to conduct information audit to demonstrate GDPR compliance}}& {{HI is required to keep an up-to-date and detailed list of their processing activities using a data protection impact  assessment. The list should include the purposes of the processing, what kind of data you process and who has access to it in the organization}}
         \\
        \hline
        
        \textbf{Art. 6 (Lawfulness of processing), Art. 7 (Conditions for consent)
        } & {Legal justification for processing  health data}& {HI can justify the purpose according to one of the six conditions. E.g Patients has given consent for the processing. Extra obligation such as the opportunity to revoke consent must be available to patients} \\
        \hline
        
         \textbf{Art. 12 (Transparent information, communication and modalities for the exercise of the rights of the data subject)} & {Clear information about the data processing and legal justification in privacy policy}& {HI is obligated to inform patients that health data is collected. HI should explain why this is collected, how it is processed, who has the access and how it is secured using clear and plain language, particularly when addressing specifically to a child.} \\
        \hline
        
        \textbf{Art. 33 (Notification of a personal data breach to the supervisory authority),
        Art. 34 GDPR (Communication of a personal data breach to the data subject)
        } & {Have a process to notify the authorities in the event of a data breach}& {HI is required to notify the supervisor authority in their jurisdiction within 72 hours learning of the health data breached or exposed.  Patients should be notified without undue delay in plain language, if the breach is likely to put them at risk.} \\
        \hline
        
        \textbf{Art. 32 (Security of processing)} & {Encrypt, pseudonymize or anonymize personal data whenever possible} & {HI is to encrypt, pseudonymize or anonymize PHI whenever feasible.} \\
        \hline
        
        \textbf{Art. 25 (Data protection by design and by default), Art. 5 (Principles relating to processing of personal data)} & {Data  protection is considered at all times, including at the beginning of developing a product}& {HI should implement appropriate technical (encryption) and organizational measures(deleting patients data that is no longer needed) to protect data. HI which adheres to data protection principles when processing of personal data is involved.} \\
        \hline
        
        \textbf{Art. 25 (Data protection by design and by default) } & {Designated person for ensuring GDPR compliance across the organization }& {HI should designate someone that is accountable for GDPR compliance which includes evaluation of data protection policies and the implementation of policies. HI should be able to verify the patient's identity.} \\
        \hline
        
        \textbf{Art. 15 (Right of access by the data subject)}&  {Able to verify the patients' identity} & {HI is obligated to furnish patients with complete information when they request it and  should be able to comply within a month. HI should be able to verify the patient's identity. } \\
        \hline
        
        \textbf{Art. 17 (Right to erasure/ ‘right to be forgotten’)} & {Easy to delete personal data upon request} & {Patients should have the right to request to delete all health data and HI should honour their request within a month. HI may have grounds to deny the request  such as compliance with a legal obligation. HI should be able to verify the patient's identity.} \\
        \hline
        
        \textbf{Art. 18 (Right to restriction of processing)} & {Easy to stop data processing upon request} & {Patients can request HI to restrict or stop processing their health data if certain grounds apply, such as dispute about the lawfulness of the processing. HI may be allowed to keep storing their data although the processing is restricted.}\\
        \hline
        
        \textbf{Art. 24 (Responsibility of the controller)} & {Establish the responsibility and liability of the controller} & {Any processing of personal data carried out by HI or on HI’s behalf, responsibilities should be established which includes implementing appropriate technical and organisational measures. This is to ensure and to be able to demonstrate that processing is performed lawfully}\\
        \hline
        
        \textbf{Art. 20 (Right to data portability)} & {Easy to receive a copy of your personal data and share with another in a simple format} & {From a privacy standpoint, GDPR offers higher patients autonomy over their data, instead of HI. This means patients should be able to receive health data in a readable format or share with other HI.}\\
        \noalign{\hrule height 2pt}
    \end{tabular}
    \end{center}
\end{table*}

Blockchain structure can enhance the traceability of data making transactions auditable and transparent. In addition, storing data on the blockchain ledger could increase data integrity due to the inherited immutability property. However, any form of information stored on blockchain remains on blockchain and it might be violating GDPR since patients should have the right to erase their personal data. Although, it is common that national health data law prohibits the deletion of patient data from medical health records.  Storing metadata could be an alternative to storing the full dataset. Storing metadata on the blockchain can pseudonymize a patient's identity and to further protect patient's identity, encryption technology such as zero-knowledge proofs could be implemented to prevent any forms of identification \cite{8865045}.

Table \ref{table2} presents a summary of how four different blockchain-based applications (presented under Section IV) comply with the relevant GDPR articles presented in Table \ref{table1}. 

\begin{table*}[h]
\centering
    \caption{Comparative analyzes}\label{table2}
        \begin{center}
        \begin{tabular}{|L{4.5cm}||L{4.5cm}|L{1.5cm}|L{1.5cm}|L{1.5cm}|L{1.5cm}|}
        \noalign{\hrule height 2pt}
        
       \multirow{2}{*}{\parbox{4.5cm}{\centering Feature}} & \multirow{2}{*}{\textbf {GDPR article}} & \multicolumn{4}{c|}{Blockchain application} \\
        \cline{3-6}
        & & \textbf{MedRec} & \textbf{EMRshare} & \textbf{FHIRchain} & \textbf{VerifyMed} \\
        \noalign{\hrule height 2pt} 
        \multirow{2}{*}{\parbox{4.5cm}{\textbf{Able to conduct information audit to demonstrate GDPR compliance}}} & \textbf{Art. 30 GDPR (Records of processing activities)} & \textbf{Yes} &\textbf{Yes} &\textbf{Yes} &\textbf{Yes} \\
        \cline{2-6}
        
          & \textbf{Art. 35 GDPR (Data protection impact assessment)} & \textbf{N/A} &\textbf{N/A} &\textbf{N/A} &\textbf{N/A} \\
        \noalign{\hrule height 1pt}
        
        \multirow{2}{*}{\parbox{4.5cm}{\textbf{Legal justification for processing  health data}}} & \textbf{Art. 6 GDPR (Lawfulness of processing))} & \textbf{N/A} &\textbf{N/A} &\textbf{Yes} &\textbf{N/A} \\
        \cline{2-6}
        
        \textbf{} & \textbf{Art. 7 GDPR (Conditions for consent)} & \textbf{N/A} &\textbf{Yes} &\textbf{Yes} &\textbf{N/A} \\
        \noalign{\hrule height 1pt}
        
        \textbf{Clear information about the data processing and legal justification in privacy policy} & \textbf{Art. 12 GDPR (Transparent information, communication and modalities for the exercise of the rights of the data subject)} & \textbf{Yes}& \textbf{Yes}& \textbf{Yes}& \textbf{Yes}\\
        \noalign{\hrule height 1pt}

        \multirow{2}{*}{\parbox{4.5cm}{\textbf{Have a process to notify the authorities in the event of a data breach}}} & \textbf{Art. 33 GDPR (Notification of a personal data breach to the supervisory authority)} & \textbf{No}& \textbf{No}& \textbf{No}& \textbf{No}\\
        \cline{2-6}
        
        \textbf{} & \textbf{Art. 34 GDPR (Communication of a personal data breach to the data subject)} & \textbf{No}& \textbf{No}& \textbf{No}& \textbf{No}\\
        \noalign{\hrule height 1pt}
        
        \multirow{2}{*}{\parbox{4.5cm}{\textbf{Data  protection is considered at all times, including at the beginning of developing a product}}} & \textbf{Art. 25 GDPR (Data protection by design and by default)} & \textbf{Yes}& \textbf{Yes}& \textbf{Yes}& \textbf{Yes}\\
        \cline{2-6}
        
        \textbf{} & \textbf{Art. 5 GDPR (Principles relating to processing of personal data)} & \textbf{N/A}& \textbf{N/A}& \textbf{N/A}& \textbf{N/A}\\
        \noalign{\hrule height 1pt}
        
        \textbf{Encrypt, pseudonymize or anonymize personal data whenever possible} & \textbf{Art. 32 GDPR (Security of processing)} & \textbf{Yes}& \textbf{Yes}& \textbf{Yes}& \textbf{Yes}\\
        \noalign{\hrule height 1pt}
        
        \textbf{Designated person for ensuring GDPR compliance across the organization} & \textbf{Art. 25 GDPR (Data protection by design and by default)} & \textbf{No}& \textbf{No}& \textbf{No}& \textbf{No}\\
        \noalign{\hrule height 1pt}
        
        \textbf{Should be able to verify the patients identity.} & \textbf{Art. 15 GDPR (Right of access by the data subject)} & \textbf{Yes}& \textbf{Yes}& \textbf{Yes}& \textbf{No}\\
        \noalign{\hrule height 1pt}
        
        \textbf{Easy to delete personal data upon request} & \textbf{Art. 17 GDPR (Right to erasure/ ‘right to be forgotten’)} & \textbf{No}& \textbf{No}& \textbf{No}& \textbf{No}\\
        \noalign{\hrule height 1pt}
        
        \textbf{Easy to stop data processing upon request} & \textbf{Art. 18 GDPR (Right to restriction of processing)} & \textbf{N/A}& \textbf{No}& \textbf{N/A}& \textbf{No}\\
        \noalign{\hrule height 1pt}
        
        \textbf{Establish the responsibility and liability of the controller} & \textbf{Art. 24 GDPR (Responsibility of the controller)} & \textbf{No}& \textbf{No}& \textbf{No}& \textbf{No}\\
        \noalign{\hrule height 1pt}
        
         \textbf{Easy to receive a copy of your personal data and share with another in a simple format} & \textbf{Art. 20 GDPR (Right to data portability)} & \textbf{Yes}& \textbf{Yes}& \textbf{Yes}& \textbf{No}\\
        \noalign{\hrule height 2pt}
    \end{tabular}
    \end{center}
\end{table*}

All four concepts use slightly different components of blockchain technologies. For example, FHIRchain uses a public blockchain (Ethereum) while EMRShare uses a permissioned blockchain (Hyperledger). These blockchains vary in some properties, such as the degree of visibility, but both types of blockchains can store transaction chronological with high data integrity due to the immutable structure.  Blockchain data structure is easily auditable, which can make it inherited compliant with Art 30 and 35. 

Identity management also forms a core technology in all these frameworks. This is one of the key compliance to Art 15. "Right access by the right data", before executing requests from patients to obtain health information or stop processing their health data. This is to prevent any misuse of private health data by the wrong person. For example, FHIRChain adopts digital identity to verify and authenticate the identity of clinicians. VerifyMed does not incorporate identity management to authenticate as the application utilizes evidence of authority to proof the credential of the clinicians. 

Table II highlights that the blockchain blockchain concepts identified for this work did not fulfill the requirement of the right to forgetting (Art. 17). Patients can have the right to request for deletion of their information but the immutability nature of blockchain contradicts this article. To circumvent this, proposed concepts, such as FHIRchain, only stores metadata and protected with encryptions. Although it is not erasure, it prevents an unauthorized person from obtaining information and linking the pseudonymized metadata to patient's identity. 
Currently, the four explored concepts did not state any procedures to notify authorities if any violation of GDPR is detected and might therefore lack compliance with (Art 33). A smart contract can be designed by sending a notification to relevant authorities when a breach is detected. In addition to that, a designated person for ensuring GDPR compliance within the network should be considered for future work.

\section{Discussion}
There is an increased focus on blockchain technology in healthcare sector in both academic spheres and the private sector with the expectation that this technology could have a positive impact on achieving better interoperability and access to health data \cite{hasselgren2020blockchain}. This can bring medical advances, such as enabling collaborative treatment and care decision. However, storing patient's health data or even metadata is considered highly sensitive and could violate patient's data privacy. In order to protect patient's health data, GDPR has defined rules and guidelines to ensure data processing and handling comply. However, research focusing on  the degree of compliance of proposed blockchain solutions to GDPR in the healthcare sector remains limited. 

The contribution of this paper explores how four different blockchain-based healthcare applications comply with the identified articles in GDPR. This analysis can provide further research guidance on how to achieve GDPR compliance and what architectural design choices that need to be considered.

As outlined under Section VI, compliance with Art 30 and 35 are achieved in the four healthcare applications identified due to the inherited characteristics of blockchain - storage of transaction chronological with high data integrity. Identity management is a core technology for healthcare application and it is also a key compliance factor in GDPR with Art 15: Right access by the right data. This is mainly to prevent any misuse of private health data by the wrong person and compliance is achieved with three out of the four concepts. 

Currently, none of the proposed concepts did fulfill the requirement of the right to be forgotten (Art 17), as shown in Table II. This indicates that patients should have the right to request the deletion of their information. However, this is often regulated by national health data laws that prohibit the deletion of data from medical health records. Nevertheless, compliance with this article is problematic due to the immutable nature of blockchain. Compliance can be achieved by making all data stored on the ledger entirely anonymous or fully encrypted. Hence, we encourage researchers to explore full anonymity in blockchain applications for this domain. 

 None of the investigate blockchain concepts did consider the process of notifying authorities if any violation of GDPR is detected stated in Art. 33 as shown in Table \ref{table2}. This could potentially be implemented by a smart contract to ensure automated and imitate notifications to relevant authorities upon data breach. A way to ensure that any new blockchain solutions that handle sensitive health data comply with GDPR, is to keep an up-to-date lists using Data Protection Impact Assessment (DPIA)(Art. 35) to any authorities or regulators upon requests. This can avoid any solution providers from subjecting to severe penalties (fines of up to 20 million dollars or 4 percent of annual revenue whichever is higher \cite{EUdataregulations2018}) and losing trusts from its users. Therefore, researchers should ensure Art. 33 and 35 are in place before deployment in the real-world scenario. 

\subsection{Conclusion}
Blockchain compliance with GDPR for healthcare applications is highly dependent on how the technology is utilized and the architectural design. It seems infeasible to conclude that specific blockchain frameworks or main blockchain characteristics are more compliant than others, it is rather use-case dependent and based on several design aspects that together could build up towards GDPR compliance. This research shows that blockchain may enhance GDPR in some aspects and be challenging with some others. It is important that this topic is being addressed and highlight potential compliance issues to increase adoption and acceptance of the technology in this field. There is no such thing as GDPR-compliant blockchain technology for healthcare, but it might be GDPR-compliant use cases and applications. We encourage future work to address GDPR compliance to get closer to real-world adoption of blockchain technology in the healthcare sector.

\bibliographystyle{unsrt}
\bibliography{bibtex.bib}

\end{document}